# FPGA-based digitizer for BGO-based time-of-flight PET

**Daehee Lee and Sun Il Kwon***

Department of Biomedical Engineering, University of California, Davis, One Shields Avenue, Davis, CA 95616, USA

E-mail: sunkwon@ucdavis.edu

## Abstract

We present a novel FPGA-based bismuth germanate (BGO) time-of-flight (TOF) digitizer, implemented on an FPGA (VC707 evaluation kit, Xilinx). This digitizer was designed to address the recently highlighted characteristics of BGO, which generates scintillation and prompt Cerenkov photons when 511-keV photon interacts with BGO. The developed digitizer independently processes these two types of photons for precise energy and timing measurements. The digitizer incorporates a noise-resistant binary counter that measures energy signals using a time-over-threshold (TOT) method. For timing measurement, we employ an embedded dual-side monitoring time-to-digital converter (TDC), which efficiently captures timing information while maintaining low resource usage. We validated the efficacy of our FPGA-based TOF digitizer through extensive experiments, including both an electrical setup and a coincidence test using BGO pixels. Our evaluations of TOT energy and timing performance utilized two 3 × 3 × 20 mm$^3$ BGO pixels coupled to CHK-HD MT silicon photomultipliers (SiPMs). The digitizer achieved a coincidence timing resolution (CTR) of 407 ps full width at half maximum (FWHM) for coincidence events falling within the full width at tenth maximum (FWTM) of the measured TOT energy spectrum. Notably, when measured with an oscilloscope, the same detector pair exhibited a CTR of 403 ps FWHM, confirming that the performance of the developed digitizer is comparable to that of oscilloscopes. With the low resource usage of our design, it offers significant potential for scalability, making it particularly promising for multi-channel BGO-based PET systems.

## 1. Introduction

Bismuth germanate (BGO) initially served as a major scintillator in the formative stages of positron emission tomography (PET) detectors, and is known for its excellent attenuation properties (Jones and Townsend 2017, Vandenberghe *et al* 2020, Yu *et al* 2022). Its widespread use underscored its importance within early PET systems. As the field advanced in the 2000s, the landscape of PET scintillators underwent a substantial transformation following the emergence of lutetium-based scintillators, including lutetium orthosilicate (LSO) and lutetium yttrium orthosilicate (LYSO) (Vandenberghe *et al* 2020, Jones and Townsend 2017). Given their superior light yield and shorter decay time, these LSO and LYSO scintillators gradually superseded BGO (Yu *et al* 2022).

As PET detector development has advanced, there has been growing interest in improving the signal-to-noise ratio (SNR) of reconstructed PET images. This objective has been pursued through sophisticated methods to measure the arrival time difference between two back-to-back 511 keV photons within the sub-nanosecond range, enabled by improvements in signal measurement techniques. These developments led to the emergence of time-of-flight (TOF) PET, with lutetium-based scintillators becoming dominant in most commercial TOF PET systems due to their superior timing performance (Surti and Karp 2020). However, the higher cost of these materials compared to BGO has inevitably increased the overall expenses associated with PET system development (Rocio A. Ramirez 2005, Yu *et al* 2022).

Recent studies on BGO, which is about three times cheaper than conventional TOF PET scintillators, reveal that approximately 17 prompt Cerenkov photons are generated when a 511-keV photon interacts with a BGO pixel, preceding scintillation photon emission (Brunner and Schaart 2017, Gundacker *et al* 2020). These Cerenkov photons are predominantly in the blue/UV range compared to scintillation photons (Kwon *et al* 2016, Trigila *et al* 2022, Kwon *et al* 2019). The significant enhancement in photon detection efficiency (PDE) of silicon photomultipliers (SiPMs), particularly in the blue/UV range, has enabled the successful measurement of these few prompt Cerenkov photons in BGO (Kratochwil *et al* 2020, Kwon *et al* 2016). Notably, utilizing scintillation photons for energy measurement and prompt Cerenkov photons for timing measurement has substantially improved the coincidence timing resolution (CTR) to less than 500 ps (Kratochwil *et al* 2021, Cates and Levin 2019, Kwon *et al* 2019). These improved CTRs have opened up possibilities for developing TOF PET detectors using cost-effective BGO instead of the more expensive LYSO or LSO materials.

However, conventional integrated DAQ digitizer systems for TOF PET applications have been developed and optimized to acquire energy and timing information solely from the large number of scintillation photons produced by scintillators such as LYSO and LSO. (Ullah *et al* 2016, Lu *et al* 2019, Son *et al* 2016, Torres *et al* 2013). These systems focus on processing large numbers of scintillation photons, making them less suitable for emerging BGO-based TOF PET systems, which require independent paths for energy measurement from scintillation and precise timing measurement from the small number of Cerenkov photons, which produce significantly weaker signals (Piller *et al* 2024). To address these limitations, we conducted this study and developed an innovative approach that independently uses Cerenkov photons for timing measurements and scintillation photons for energy measurement.

In this research, we present a BGO TOF digitizer implemented on a field-programmable gate array (FPGA), which offers rapid design capabilities and reprogrammable circuit flexibility (Sohal *et al* 2014, Machado *et al* 2019a). This study examines each critical component for energy and timing measurement, and introduces the comprehensive system developed for BGO-based TOF PET modules. We also describe an experimental setup used for extensive electrical testing and coincidence measurement. Finally, we present test results obtained using 3 × 3 × 20 mm$^3$ BGO pixels coupled to CHK-HD MT SiPMs (FBK, Italy). CHK-HD MT SiPM was chosen for its high detection efficiency in the Cerenkov wavelength range and its ability to operate at higher bias voltages, improving timing resolution. (Merzi *et al* 2023, Gundacker *et al* 2023).

## 2. Material and methods

### *2.1 Energy and Timing Signals from BGO Pixels*

The BGO pixels were coupled to CHK-HD MT SiPMs and pole-zero cancellation circuits, which generated the energy and timing signals (Gala et al 2011, Gundacker et al 2023). To verify the signal properties, we digitized signals using an oscilloscope (MSO64B, Tektronix) with a bandwidth of 4 GHz and a sampling rate of 25 GS/s. Both CHK-HD MT SiPMs, with a breakdown voltage of 32.5 V, were subjected to a bias of 49 V throughout the experiments, except during a bias optimization test in *Section IV. D*. Fig. 1 presents typical energy and timing signals acquired using two 3 × 3 × 20 mm$^3$ BGO pixels (Epic Crystal, China) and a $^{22}$Na point source. They are presented here preemptively as they directly illustrate the key signal characteristics that support the proposed method. Notably, Fig. 1.(b) clearly shows spike-shaped signals during energy measurements, produced by the fast signal component of a single fired cell in a SiPM (Merzi *et al* 2023, Gundacker *et al* 2023). Depending on the measurement method, particularly in the time-over-threshold (TOT) method, these spike signals may cause early termination of the energy measurement process, making accurate energy measurement difficult. This issue is discussed in more detail in the following section. The energy signal is slower than the timing signal is due to the offset time delay caused by the difference in amplifiers for generating the energy signal and the timing signal.

### *2.2 Energy Measurement Unit*

The proposed FPGA-based digitizer employed the TOT technique as an energy measurement unit due to its efficient use of the FPGA's inherent digital components: a binary counter (BC) and a comparator (Won *et al* 2021, 2020). The TOT method, common in particle physics and imaging fields, allows for the transformation of energies into digital values in a resource-effective way (Jakubek 2011, Chang et al 2017, Urban et al 2023).

The TOT method involves establishing a predetermined threshold for one of the comparator's inputs. The subsequent BC quantifies the duration in which the energy signal, acting as the second input to the comparator, surpasses this threshold. The TOT counter value at the BC is incremented in synchronization with a system clock, thereby signifying the energy of the input signal. For instance, with a system clock of 550 MHz (1.8 ns clock period), an input energy signal with a duration of 4 ns is translated into a counter value of #2, representing an energy signal range from 3.6 ns (1.8 ns × #2) to 5.4 ns (1.8 ns × #3). Although a faster clock frequency enables more precise TOT energy measurements, power consumption also increases significantly. Therefore, the system clock frequency should be optimized for each dedicated system (Drozd *et al* 2021). In addition, the recorded TOT values may exhibit distortions in the high energy region depending on the threshold level, indicating the necessity for optimal threshold voltage selection for the TOT spectrum (Jakůbek 2009, Jakubek 2011).

Managing local spikes in energy signals with the conventional TOT method, as depicted in Fig. 1.(b), presents a significant challenge for the BC. Local spikes on the rising phase of the energy signal could cause multiple transitions at the comparator's output, potentially prematurely terminating the BC operation and recording multiple lower TOT values instead of the correct higher TOT energy value. To address this issue, we proposed a noise-resistant BC (NRBC) incorporating a double-check logic (DCL). The DCL updates the comparator's output on a user-defined period (UPT) time basis. By filtering out the noise (jitters caused by local spikes), the BC maintains stability. Consequently, the NRBC enables the prevention of incorrect measurements or premature terminations caused by local spikes, ensuring accurate TOT measurements. The NRBC design is illustrated in Fig. 2.(a).

### *2.3 Timing Measurement Unit*

The time-to-digital converter (TDC) is a key component in measuring arrival-time differences of coincident events between two BGO pixels. FPGA-based TDCs primarily utilize delay lines or multiple delay lines (CARRY4 chains) for picosecond-scale time measurements, conventionally focusing solely on the start of propagation (SOP) signal through the delay line (Kim *et al* 2023, Szplet and Czuba 2021, Wang *et al* 2018, Qin *et al* 2020). These delay lines in TDC convert time into a digital code, linearly matching input times to the SOP status. However, the propagation depends on power, voltage, and temperature (PVT) conditions. As such, variations in PVT can deteriorate accurate time measurement (Machado *et al* 2019b). This requires additional resource-consuming circuitry for the TDC to correct the measured time (Won *et al* 2016, Machado *et al* 2019b). The elevated demand for resources for TDC performance presents a challenge in a complete BGO TOF PET system designed for clinical use, which requires a substantial number of channels to create a gantry that fully encapsulates the patient. Given the restricted resources available for circuit implementation in the selected FPGA chip, minimizing the physical resource usage for each TDC implementation is crucial.

To address this concern, we have incorporated a dual-side monitoring (DSM) TDC scheme in this study. Unlike traditional TDCs, the DSM TDC monitors both the SOP signal and the end of propagation (EOP) for each time conversion. This strategy allows the DSM TDC to apply an effective correction to the SOP using the EOP. As a result, it maintains stable time measurement characteristics under PVT variations, eliminating the need for additional extensive correction circuits. This approach presents a resource-efficient solution for TDC implementation in the FPGA while preserving accurate time measurement. Fig. 2.(a) illustrates the block diagram of the DSM TDC employed for the BGO TOF digitizer, detailing the operational procedures of both a main controller and a channel controller in Fig. 2.(b). The operation of the channel controller and the main controller is described in *Section D*.

Furthermore, we employed the cross-detection sampling method, which alternates adjacent odd and even outputs at each CARRY4 to yield bubbleless thermometer codes. For instance, instead of sampling thermometer codes in the sequence of 1-2-3-4, we opt for a 2-1-4-3 order (Lee *et al* 2023). This innovative method of generating bubbleless thermometer codes increases resource efficiency by eliminating the need for a bubble correction circuit. The SOP is fine-tuned using the EOP, yielding the final absolute time information for each time measurement. The final absolute time value is then merged with the coarse counter value to produce the final TDC value (Won and Lee 2016, Di Zhang *et al* 2020). The coarse counter in the BGO TOF digitizer utilizes a 12-bit BC to ensure a broad time input range of up to approximately 7.4 μs (4096 × 1.8 ns) from each system reset.

### *2.4 Two Controllers for BGO TOF Digitizer Operation*

The channel controller in the BGO TOF digitizer initially inspects *Buffered T* (timing) and *E* (energy) signals, which are the outputs of a low voltage differential signaling (LVDS) buffer for the timing and energy inputs, respectively. The *Gated T* and *E* signals are activated when the *Buffered T* and *E* signals are stable, as controlled by the channel controller. These gated signals are subsequently used for timing and energy measurements in the DSM TDC and NRBC, respectively.

Following the gating process for the timing and energy signals, the channel controller verifies the time difference between the *Gated T* and *E* signals for each channel input. This enables the rejection of timing signals inaccurately triggered by noise. If the timing signal is detected earlier than the noise rejection window from the energy signal input, the channel controller resets its channel to repeat the signal check stage for *Buffered T* and *E*. An optimal noise rejection window is determined based on experimental results.

The channel controller further examines the measured TOT value to verify if it exceeds a predefined cut-off TOT value. This examination is essential to reject events triggered by system noise in the FPGA chip. The processed results, labeled as an energy flag, are subsequently forwarded to a main controller. The main controller then determines if each collected coincidence event from two channels is valid.

The main controller, located outside the BGO TOF digitizer modules, is responsible for managing multiple BGO TOF digitizer channels and determining whether both energy signals surpass the predetermined minimum TOT value. This information is communicated through the energy flag from each channel controller. If energy flags from both channels register as high, all recorded values (including TDC outputs, coarse counters, and TOT values) from each channel are transmitted to a computer via the universal asynchronous receiver-transmitter (UART) logic. Fig. 2.(b) illustrates the operational procedures of both the main and channel controllers.

The system clock for the BGO TOF digitizer operates at 550 MHz, with one clock cycle (1.8 ns) covering 44 CARRY4s in a delay line (43 CARRY4s for SOP and an additional CARRY4 for EOP). To optimize power consumption, the UART logic operates at a significantly lower frequency of 55 MHz (UART clock). To apply consistent effective threshold voltages for both energy and timing signals, we utilized LVDS input protocols and corrected for offset differences in these signals due to circuit variations.

## 3. Experimental setup

NRBC and DSM TDC functionalities in the BGO TOF digitizer were evaluated using an electrical setup that utilized a single output from a function generator (SDG6052X, Siglent Technologies), as shown in Fig. 3. The evaluation process consisted of two main phases: NRBC and DSM TDC functionality assessments with the electrical test setup.

For the NRBC functionality evaluation, the function generator's output was split via a T-connector, with both signals fed into the Ch1 E and Ch2 E ports on the FPGA board. The pulse width varied from the generator's minimum value (4 ns) up to the measurable counter value, #1,000 for NRBC in this study. The NRBC featured its in-built DCL and an offset minimum counter value determined by the UPT. The NRBC's maximum count value was set to #1,000, enabling a measurable count range from #(1 + UPT) to #1000.

In assessing the TDC functionality, time delays between Ch1 T and Ch2 T were introduced by copying the same signal using five different cable combinations: Cable A-B, A-C, A-A’, B-A’, and C-A’. A and A' are cables of identical length, with A' designated to differentiate between the two. These intervals were strategically chosen to validate the TDC's performance both within and beyond a single system clock period (1.8 ns) and were cross-verified using the oscilloscope. Time differences within the system clock period of 1.8 ns were generated with three different cable combinations: cables A-B, A-A’, and B-A’. Time differences that exceeded a single system clock cycle were produced using cables A-C or C-A’. The intervals, marked as differences between Ch1 T and Ch2 T, were set to 2,050 ps (Cable A-C), 260 ps (Cable A-B), 0 ps (Cable A-A’), -260 ps (Cable B-A’), and -2,050 ps (Cable C-A’).

Throughout these tests, the output of the function generator was maintained at a consistent amplitude of 1 Vpp. This resulted in an effective 500 mV applied to all negative LVDS inputs ($VT_{Energy}$ or $VT_{Timing}$), as demonstrated in Fig. 2.(a). During these experiments, the NRBC or DSM TDC modules were deactivated to assess the independent functionality of each module.

The CTR measurement involved two identical BGO pixels placed inside a dark box, each coupled with a CHK-HD MT SiPM. Both timing (Ch1 T and Ch2 T) and energy signals (Ch1 E and Ch2 E) were connected to each BGO TOF digitizer. To compensate for the offset voltage differences across energy and timing signals, individual threshold voltages were applied to the negative inputs of the LVDS buffers. For optimal CTR, the $VT_{Timing}$ was set slightly above the noise level (Kwon *et al* 2019, 2016). An effective $VT_{Timing}$ of 4 mV, determined based on experimental results, was applied across all timing channels and fixed in all subsequent experiments.

Regarding $VT_{Energy}$, an optimum UPT for NRBC operation was explored by applying BGO energy signals to the BGO TOF digitizer. The jitter at *Gated E* caused by local spikes in the rising part of the energy signals was investigated, and the optimum UPT was selected for stable TOT energy acquisition. Effective $VT_{Energy}$ values of 50 mV, 100 mV, 150 mV, 200 mV, and 250 mV were investigated to optimize the TOT energy performance for enhanced timing performance.

## 4. Results and discussion

### *4.1 Evaluation of the NRBC Functionality*

Throughout the electrical experiments, the input pulse width was measured by the NRBC, utilizing the function generator output. Pulse widths of 4, 10, 100, 200, 500, and 1800 ns were tested, while the pulse period was maintained at a constant 3 μs. Start points for measuring TOT values at the BGO TOF digitizer were randomized due to the asynchronization between the 550-MHz system clock for the BGO TOF digitizer and the input frequency from the function generator. As a result, variations were observed in the measured TOT values. For each pulse width setting, 1,000 samples were collected, and the TOT values were then averaged, as depicted in Fig. 4. For instance, the measured average TOT values were #3.8, #4.9, and #7.1 for 4 ns pulse width inputs with 1, 2, and 4 clock UPT values, respectively.

Fig. 4 shows average TOT values for six distinct input pulse widths with three different UPT conditions. As shown in an enlarged graph in Fig. 4, various UPTs resulted in offset counter values for each UPT setup. However, these offsets did not affect the linearity of the NRBC, as indicated by all calculated R-squared values exceeding 0.9999. The robust performance of the NRBC was confirmed across all UPT setups, demonstrating its superior linearity throughout the entire measurable TOT range.

### *4.2 Evaluation of the DSM TDC Functionality*

Histogram data of each measured time difference were fitted with a Gaussian curve to calculate the full width at half maximum (FWHM) value. A time bin of 10 ps was used for the histogram plot, providing a good fit across all measured points. Fig. 5 displays the FWHM values alongside corresponding time measurements. For the 0 ns (Cable A-A') input, the offset time difference between Ch1 T and Ch2 T was approximately 480 ps, due to internal routing differences from an input pad to an input logic component, which is the first component for the DSM TDC. Although this offset could be corrected by subtracting the offset time difference from the result, the original data were analyzed in the current stage to present the unmodified information. Notably, the DSM TDC accurately identified all five time-intervals, exhibiting an average FWHM of 26.37 ps. Meanwhile, the oscilloscope achieved an average FWHM of 30.12 ps for the same input data.

To evaluate the linearity of the DSM TDC, the centers of each Gaussian fit were used in the linearity calculation. The calculated linearities, as indicated by the R-squared values, were 0.9998 for the DSM TDC and 0.9889 for the oscilloscope. Thus, the DSM TDC demonstrated not only high levels of accuracy but also notable linearity. The comparatively lower performance of the oscilloscope might have resulted from its limited sampling frequency of 25 GS/s. These findings emphasize the effectiveness of the DSM TDC for precise time interval measurements.

### *4.3 TOT Energy Spectra with BGO Pixels*

An optimal UPT value is of critical importance for the NRBC, as it is directly linked to the termination of each TOT measurement at the BC by the channel controller. This termination occurs when jitters are introduced at *Gated E* due to the crossing action between local spikes and $VT_{Energy}$ at the LVDS buffer. Hence, eliminating any jitter on *Gated E* is imperative for obtaining the correct TOT energy of each photon interaction event. For practical BGO applications, an optimal UPT for the NRBC was determined based on the pulse widths of the jitter signals.

The coincidence experimental setup was employed to monitor *Gated E* signals from an actual BGO pixel. The *Gated E* signal was routed to an external oscilloscope from the FPGA. An investigation was carried out into the pulse widths of the jitter at *Gated E* with five different $VT_{Energy}$ values. Three thousand samples were collected for each $VT_{Energy}$. The measured pulse widths were histogrammed and normalized in Fig. 6, as a function of the system clock period (1.8 ns). Values at one and two clocks decreased as the $VT_{Energy}$ threshold increased due to multiple pulses at *Gated E* being triggered by local spikes. Notably, none of the jitter pulse widths exceeded 5 clocks for all $VT_{Energy}$ values. Based on these results, a decision was made to select 4 clocks as an optimal UPT value. This selection established a minimum countable TOT value of five in the NRBC for all subsequent experiments discussed in this study.

By applying the optimal 4 clocks for UPT, events surpassing a TOT value of #60 were processed by both Ch1 and Ch2 channel controllers in an attempt to further discard noise-induced events. Apart from the electrical test, the NRBC was operated with the DSM TDC for a coincidence experiment, demonstrating the full operation of the BGO TOF digitizer. For the TOT energy spectra, a total of 90,900 events for each $VT_{Energy}$ threshold were collected by the NRBC. The acquired TOT energy spectra were plotted using a bin size of 10 TOT values, as shown in Fig. 7. Owing to the applied cut-off TOT value (#60) at the channel controller, zero counts were observed below #60 TOT values in the TOT spectra. Higher values around the #60 TOT value were attributed to system noise.

Each 511 keV peak in the TOT energy spectra was fitted with a Gaussian function. No TOT energy distortion corrections were applied to the energy spectra presented in Fig. 7, resulting in the lack of a clearly defined Compton region. (Sharma *et al* 2020). Energy resolutions were calculated using FWHM values of the Gaussian fits for each case. As $VT_{Energy}$ increased from 50 mV to 250 mV, a degradation trend was observed in energy resolution, displaying 28.6%, 28.7%, 30.6%, 34.4% and 34.1%, respectively. Further increase in $VT_{Energy}$ was not indicated as the 511 keV photo peak neared the cut-off TOT (#60) value in our setup. A Compton edge wasn't distinctly visible in the TOT energy spectrum across all energy spectra. For optimum timing calculations, the data within the full width at tenth maximum (FWTM) value of each Gaussian fit for each spectrum were utilized.

### *4.4 Timing Performance with BGO pixels*

In pursuit of the optimal CTR condition, $VT_{Energy}$ thresholds for TOT energy spectra and bias voltages for CHK-HD MT SiPMs were systematically investigated, with $VT_{Timing}$ maintained at 4 mV. Fig. 8 illustrates the CTRs obtained with various $VT_{Energy}$ thresholds and bias voltages. $VT_{Energy}$ thresholds varied from 50 mV to 250 mV in 50 mV increments, while the bias voltage was adjusted from 47 V to 51 V in 1 V steps. Intriguingly, no clear correlation between energy resolution and timing performance was observed. For instance, a better energy resolution (28.6%) with a $VT_{Energy}$ of 50 mV resulted in a 440 ps FWHM in CTR, while a worse energy resolution (34.1%) with a $VT_{Energy}$ of 250 mV yielded better timing performance with a 407 ps FWHM in CTR, when a 49 V bias was applied to both CHK-HD MT SiPMs. In the experimental results, the optimal CTR was achieved with a 49 V bias voltage and a $VT_{Energy}$ threshold of 250 mV. Both FWHM and FWTM values were optimal among the 25 different setups used for CTR calculations.

Fig. 9 provides a comparison of the energy spectra and coincidence timing spectra for the BGO TOF digitizer and the oscilloscope, using the experimental setup that yielded the best CTRs for the BGO TOF digitizer. A total of 90,900 events were collected and analyzed in both cases.

The energy spectra for both systems are presented in Fig. 9.(a). The integration-based energy spectrum for the oscilloscope data is in a different unit than the TOT value for the BGO TOF digitizer, requiring the use of two different x-axes: the bottom x-axis for TOT values and the top x-axis for integration values. A Compton edge was clearly observed in the integration-based energy spectrum. The energy resolutions, calculated by Gaussian fitting of the 511 keV photopeak, were 21.39% for the oscilloscope and 34.1% for the BGO TOF digitizer. This discrepancy in energy resolution can be attributed to the differing methods used for energy spectrum calculations.

For the CTR analysis, data within the FWTM value in each energy spectrum were utilized. To verify the error value in the CTR, the experiment was conducted three times. As mentioned in *Section IV.B*, the time offset between two timing channels was about 480 ps. The centers in the CTR for both cases were aligned using two different x-axes, the bottom for the BGO TOF digitizer and the top for the oscilloscope. The BGO TOF digitizer achieved a CTR of 407 ± 8 ps FWHM, which is comparable to a CTR of 403 ± 14 ps FWHM measured using the oscilloscope. In terms of FWTM values, the BGO TOF digitizer showed slightly better performance (1215 ± 78 ps) than the oscilloscope (1309 ± 31 ps). In conclusion, the BGO TOF digitizer was successfully implemented on the FPGA platform, demonstrating CTR measurements comparable to those of the oscilloscope.

## 5. Conclusion

This study has successfully developed and demonstrated the first FPGA-based TOF digitizer specifically designed for BGO-based TOF PET systems. This novel digitizer uniquely optimizes separate measurements of energy and timing data for both scintillation and Cerenkov photons generated in BGO, marking an important advancement in PET technology to accommodate the recently highlighted improvement in timing resolution achieved by detecting Cerenkov photons.

Key achievements include the implementation of a robust, noise-resistant binary counter for BGO energy signals and a resource-efficient dual-side monitoring TDC, which achieved a superb average time bin resolution of approximately 6 ps. The digitizer exhibited effective time-over-threshold functionality for energy measurement, coupled with high accuracy and linearity in timing measurements. The TOT energy spectra are limited due to inherent distortions in the time-over-threshold (TOT) method for energy measurements, which also results in an unclear Compton region. We are preparing to implement a correction method using two comparison voltages to effectively address nonlinearity and improve the resolution of the energy spectrum (Grant and Levin 2014, Gaudin *et al* 2020). Notably, when paired with 3 × 3 × 20 $mm^3$ BGO crystals coupled to CHK-HD MT SiPMs, we achieved a CTR of 407 ps FWHM, demonstrating performance comparable to that measured using a high-performance oscilloscope.

The FPGA-based design of the BGO TOF digitizer offers significant advantages in scalability, allowing for the addition of extra channels without compromising performance or efficiency. This scalability paves the way for future advancements, with plans to increase the number of channels, enabling the control of multiple BGO-based TOF PET detector modules. This research represents an important step forward in the development of BGO-based PET scanner systems, potentially enhancing their performance and accessibility in clinical settings.

## Acknowledgements

This work was supported by National Institute of Health grants R01 EB029633. The authors thank Simon Cherry for discussions during the preparation of this work and for critically reading the manuscript. We also thank Alberto Gola, Stefano Merzi, and Michele Penna at Fondazione Bruno Kessler for providing SiPM samples and technical support.

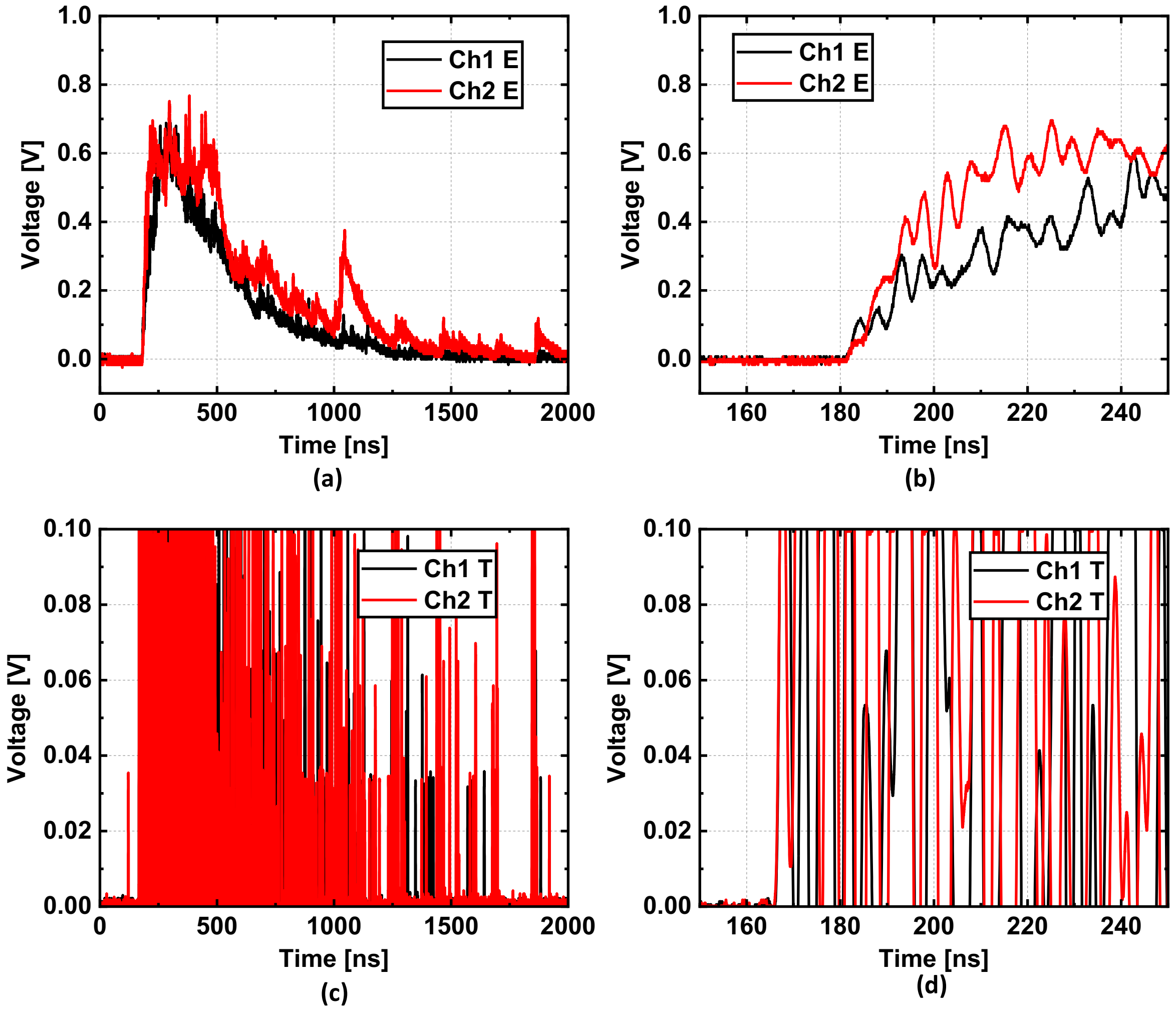

**Fig. 1.** Energy (a) and timing (c) signals of 3 × 3 × 20 $mm^3$ BGO pixels coupled to CHK-HD MT SiPMs. (b) and (d) represent the enlarged views of the energy and timing signals of the rising parts, respectively.

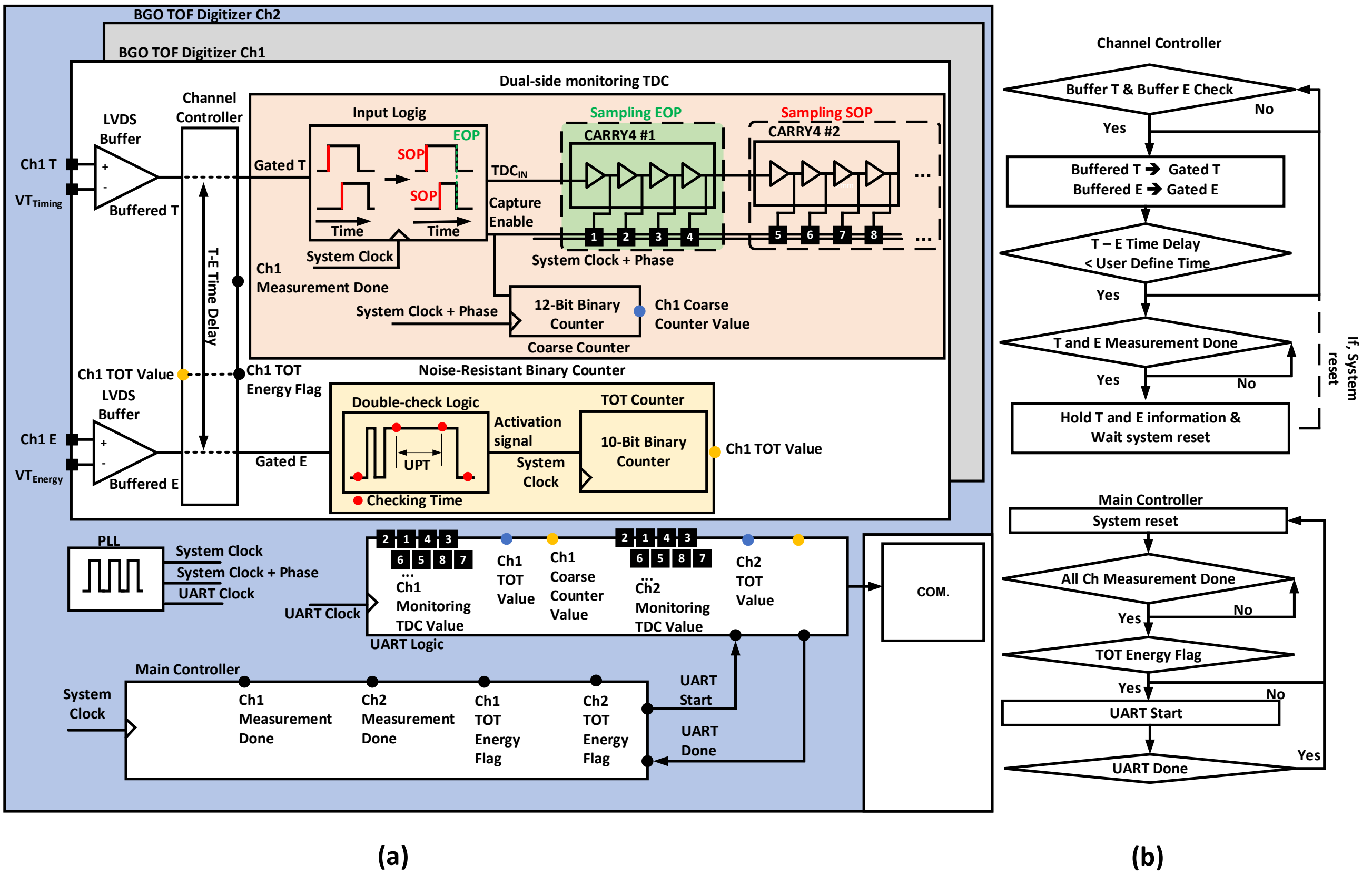

**Fig. 2.** Block diagram of the proposed BGO TOF digitizer with peripheral units (a). The BGO TOF digitizer is composed of the dual-side monitoring (DSM) TDC, the (Noise-Resistant Binary Counter) NRBC, and the channel controller used for timing and energy measurement, and to control the measurement process, respectively. Each BGO TOF digitizer channel is controlled by the main controller. operating procedures for both the channel controller and the main controller (b).

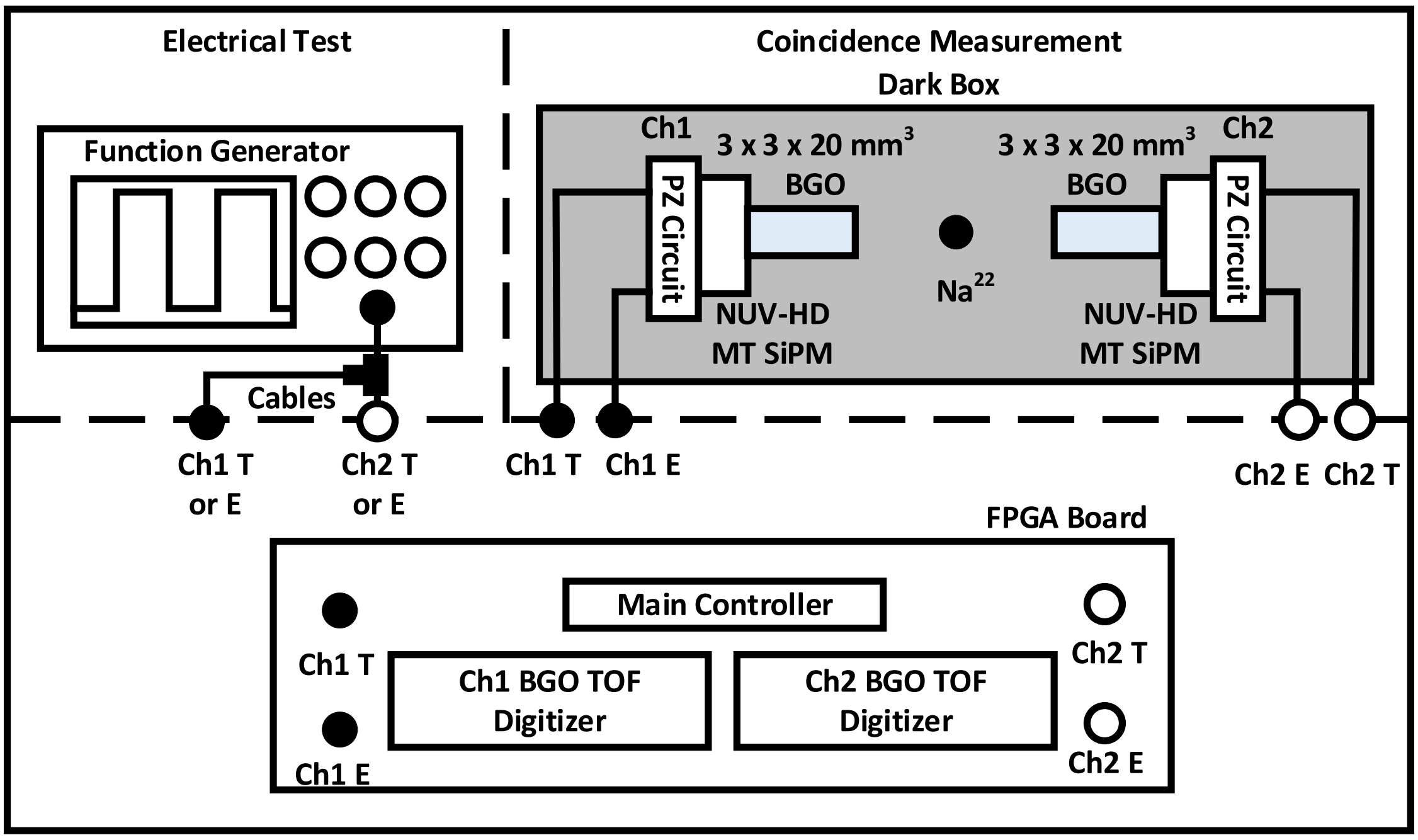

**Fig. 3.** Experimental setup for electrical tests and coincidence measurement with 3 × 3 × 20 mm³ BGO pixels.

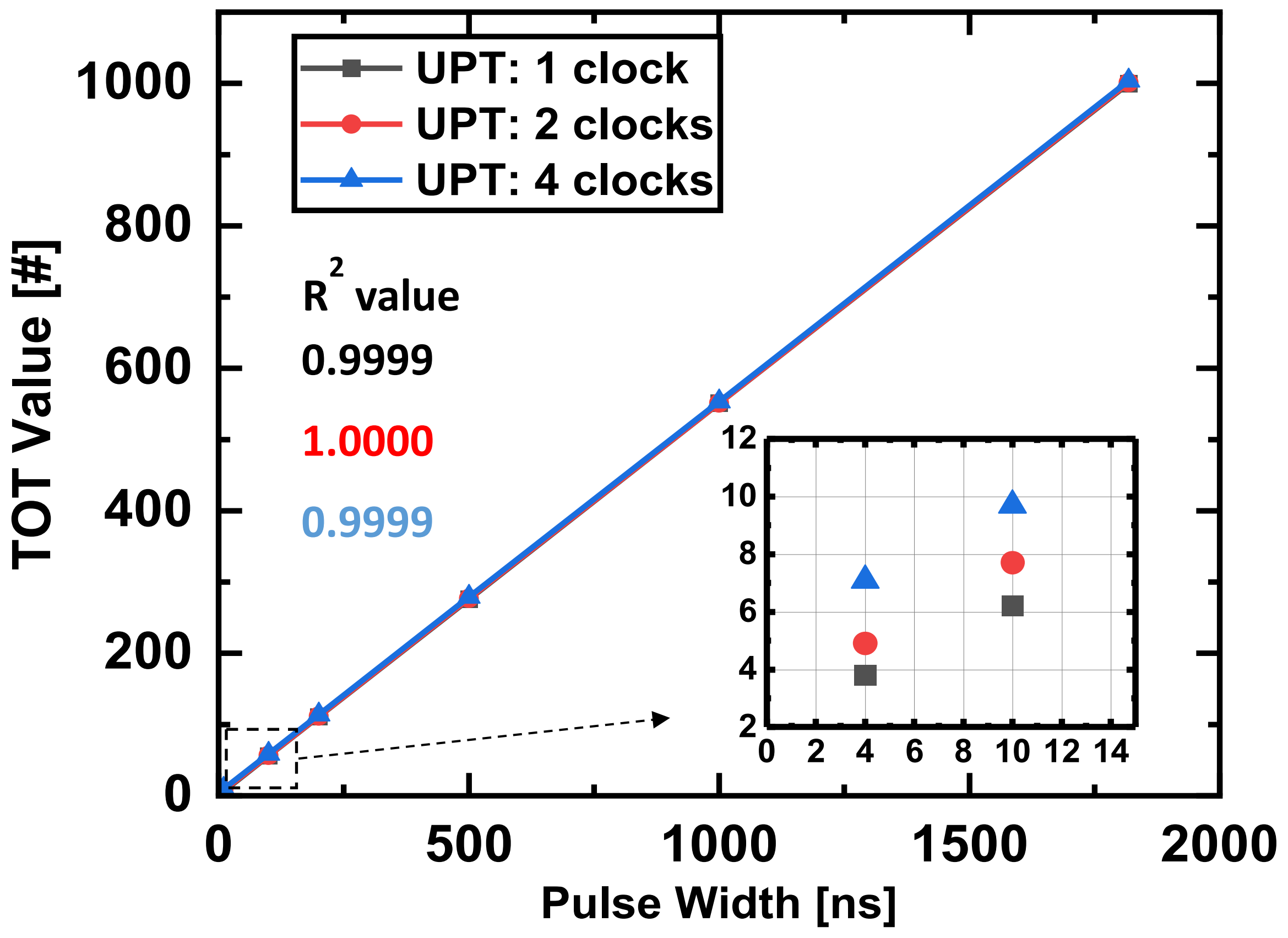

**Fig. 4.** Average TOT values measured by the NRBC using three different UPTs: 1, 2, and 4 clocks. The enlargement of the minimum countable range of the NRBC is shown in the inset.

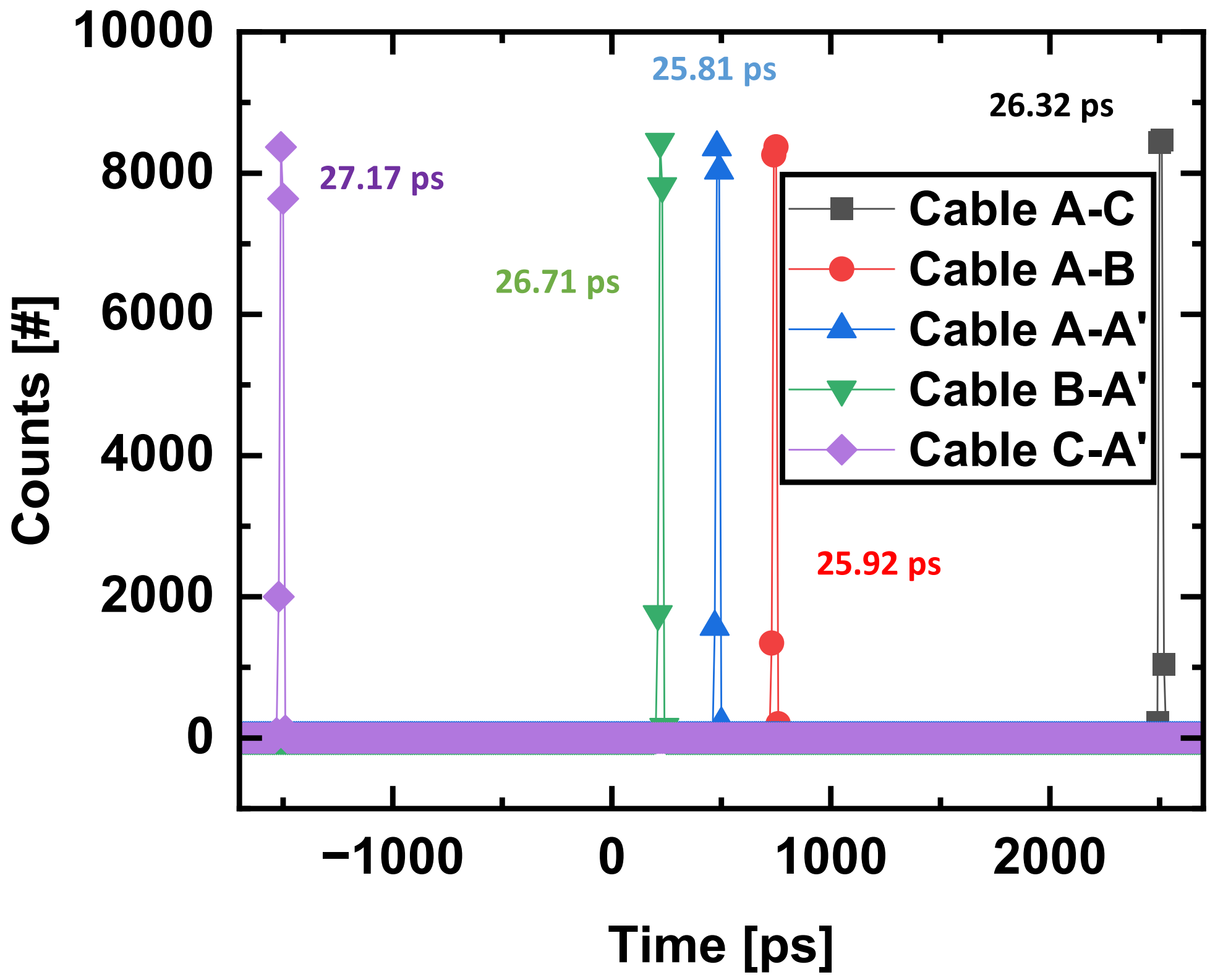

**Fig. 5.** Measured time intervals with the DSM TDC. FWHM values for each measurement are indicated.

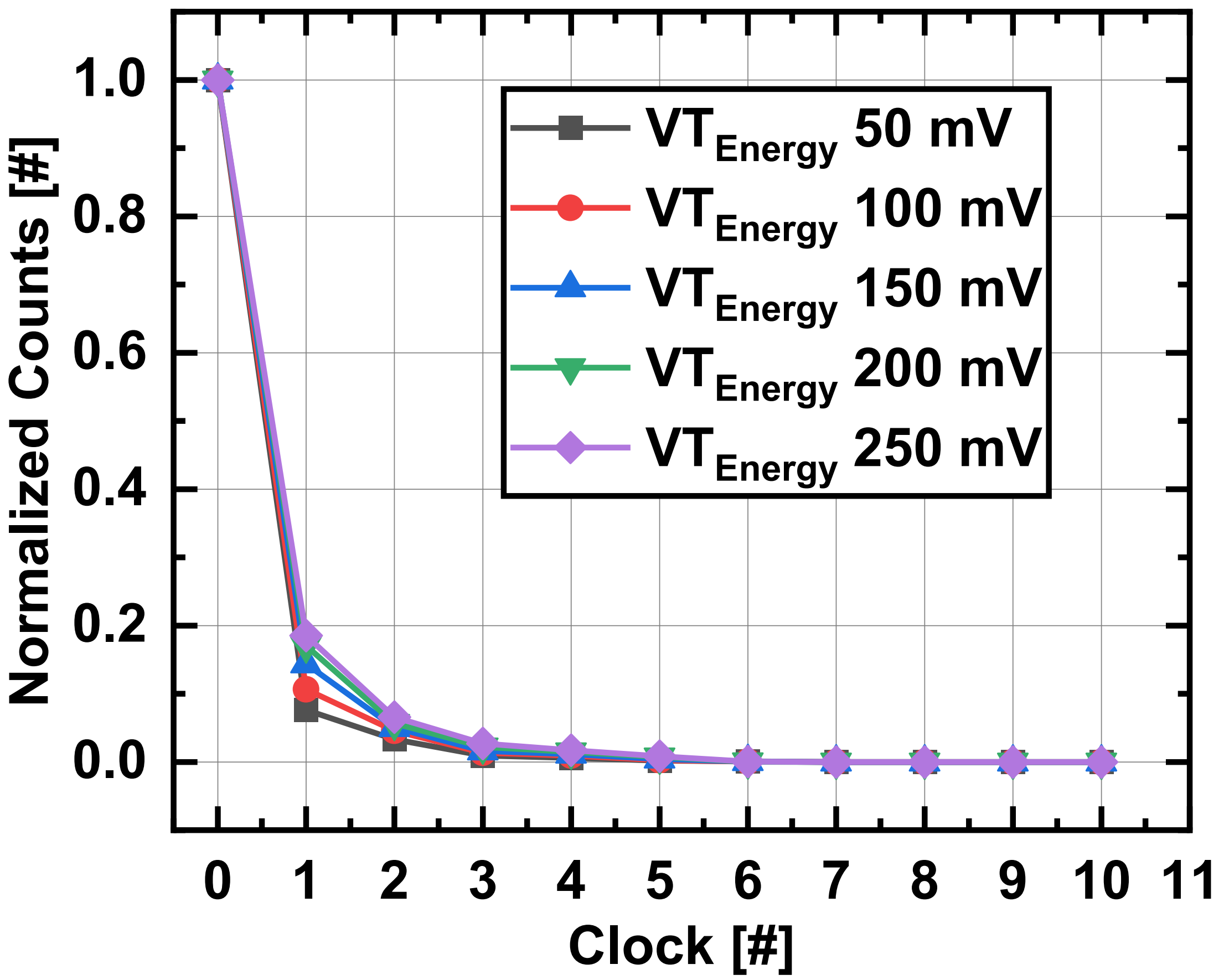

**Fig. 6.** Distribution of measured pulse widths of jitters at *Gated E* caused by local spikes.

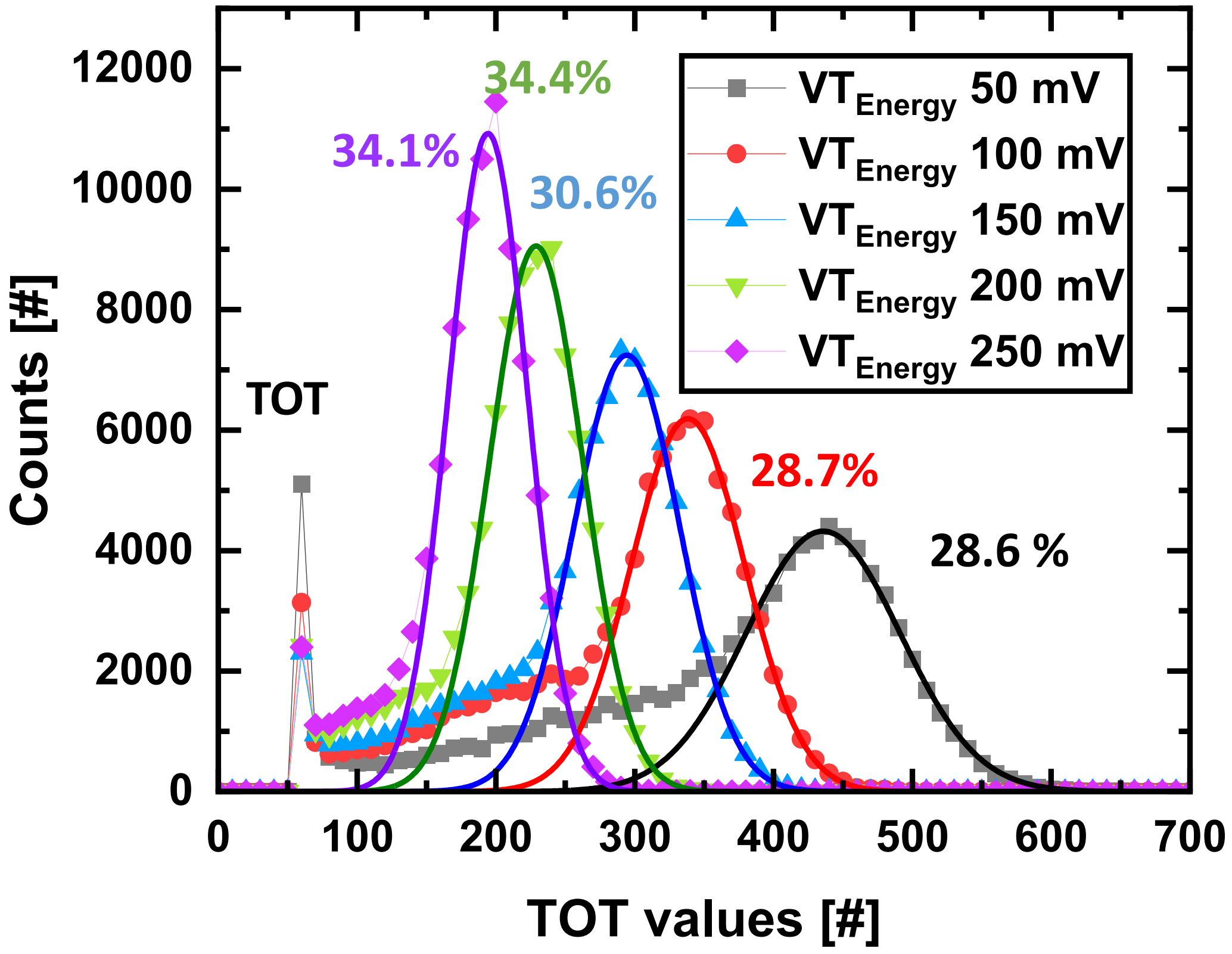

**Fig. 7.** TOT energy spectra with five different $VT_{Energy}$ thresholds, with calculated energy resolutions indicated next to each Gaussian fitting result.

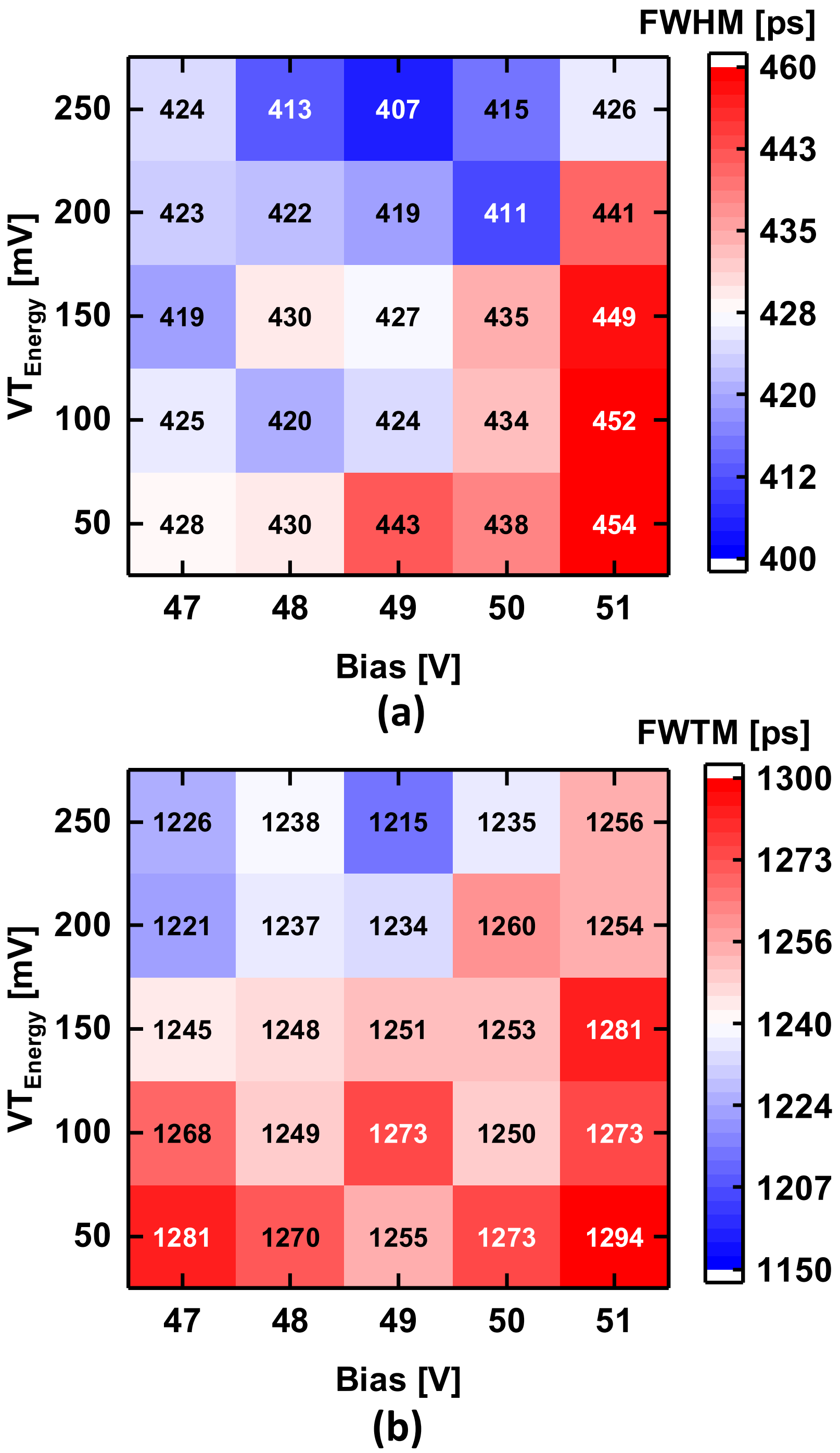

**Fig. 8.** FWHMs (a) and FWTMs (b) in CTRs for the bias and $VT_{Energy}$ sweeps measured using the developed BGO TOF digitizer

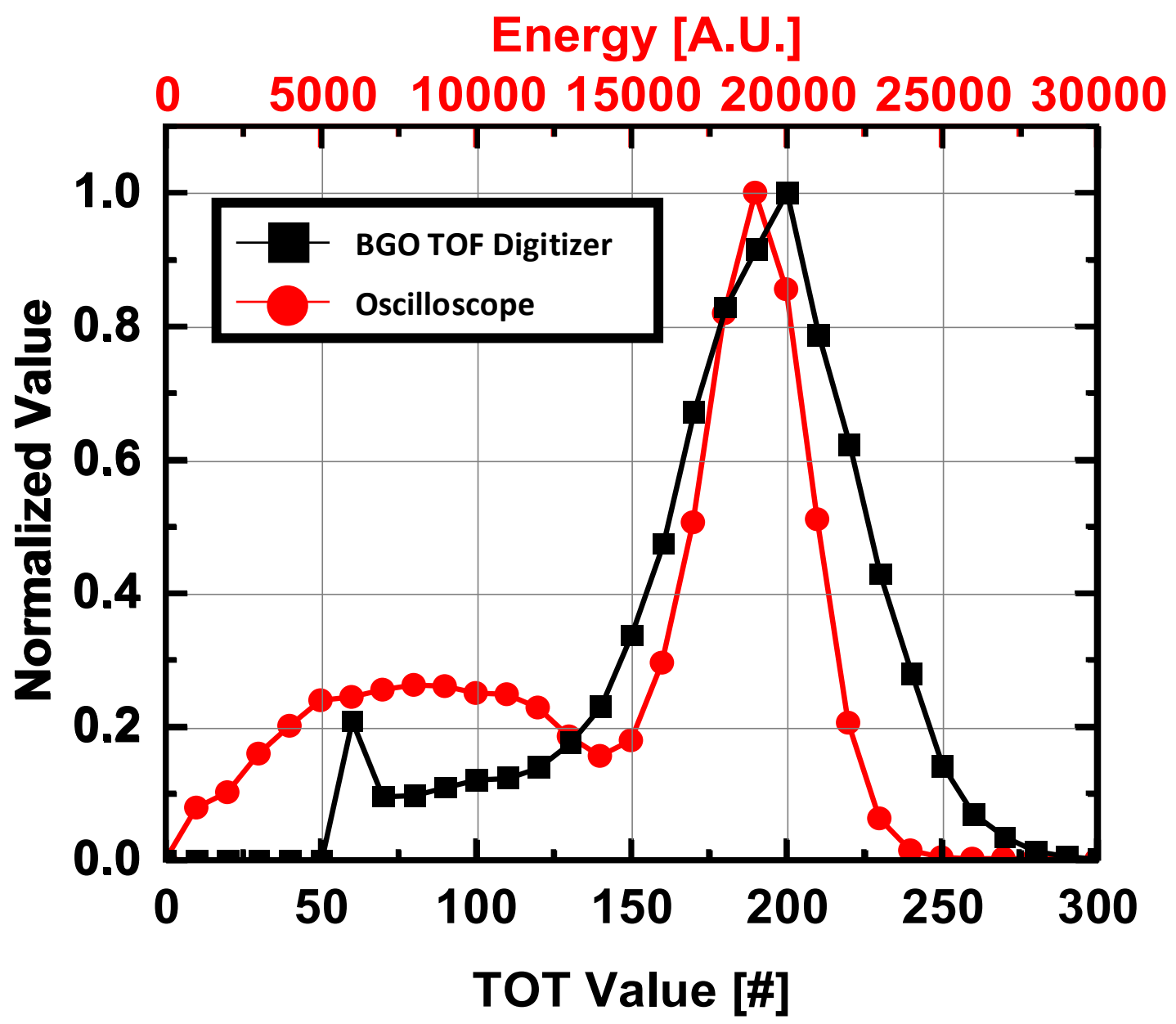

(a)

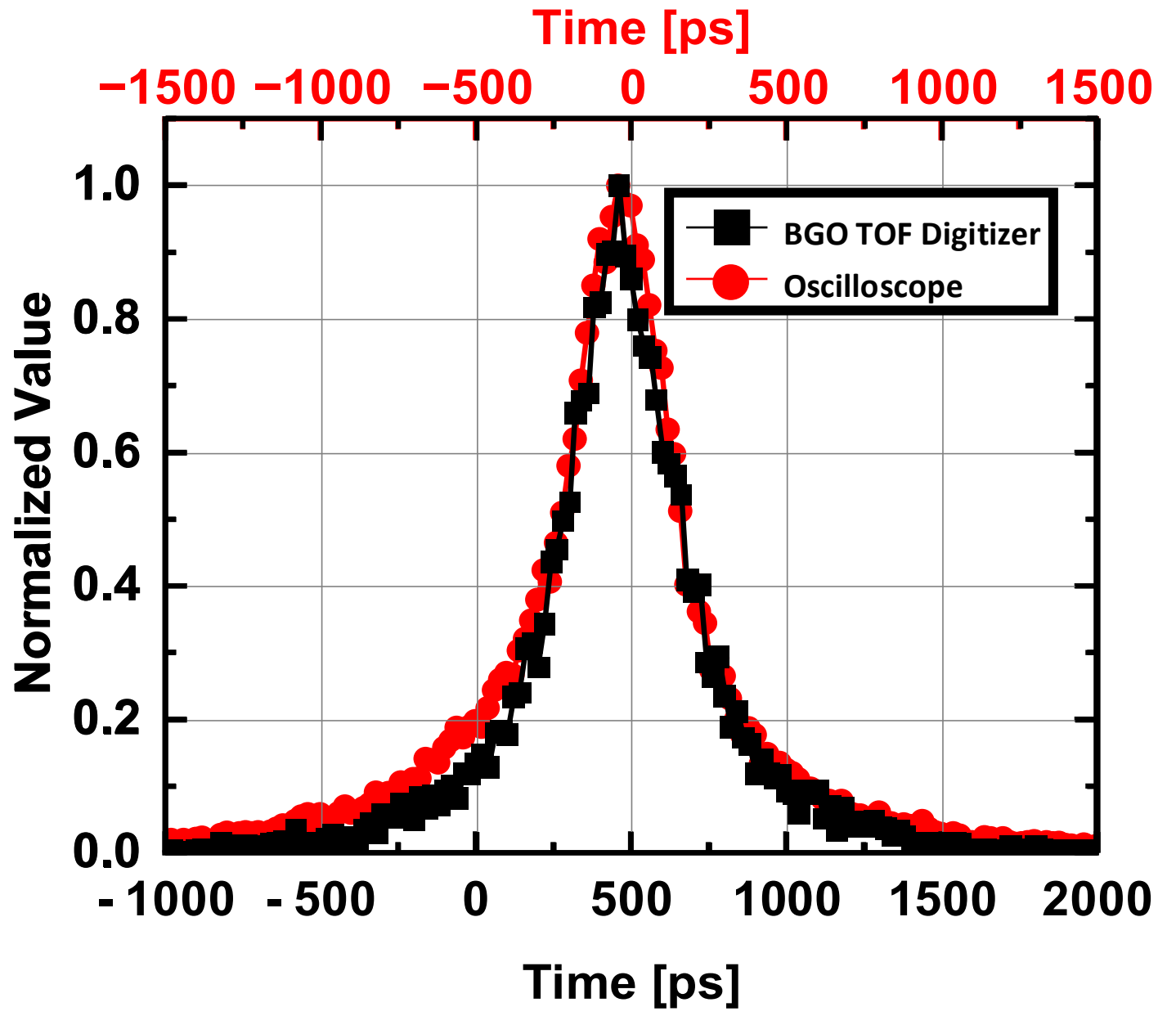

(b)

**Fig. 9.** Energy (a) and timing (b) spectra for the oscilloscope and the BGO TOF digitizer with 49 V and 250 mV for the SiPM bias and $VT_{Energy}$, respectively. The same experimental setup was utilized to minimize variation from changes in the experimental configuration.